\documentclass[epjST]{svjour}
\usepackage{graphics}
\usepackage{amsmath}
\usepackage{amssymb}
\usepackage{import}
\usepackage{bm}
\usepackage{xcolor}
\usepackage{svg}
\usepackage{appendix}
\newcommand{\rom}[1]{%
	\textup{\uppercase\expandafter{\romannumeral#1}}%
}
\begin{document}
\title{Solitary states in adaptive nonlocal oscillator networks}
\subtitle{}
\author{Rico Berner\inst{1,2}\fnmsep\thanks{\email{rico.berner@physik.tu-berlin.de}} \and Alicja Polanska\inst{1,3} \and Eckehard Sch\"oll\inst{1} \and Serhiy Yanchuk\inst{2} }
\institute{Institute of Theoretical Physics, TU Berlin, Germany \and Institute of Mathematics, TU Berlin, Germany \and Department of Physics, Imperial College, London, United Kingdom }
\abstract{
In this article, we analyze a nonlocal ring network of adaptively coupled phase oscillators. We observe a variety of frequency-synchronized states such as phase-locked, multicluster and solitary states. For an important subclass of the phase-locked solutions, the  rotating waves, we provide a rigorous stability analysis. This analysis shows a strong dependence of their stability on the coupling structure and the wavenumber which is a remarkable difference to an all-to-all coupled network. Despite the fact that solitary states have been observed in a plethora of dynamical systems, the mechanisms behind their emergence were largely unaddressed in the literature. Here, we show how solitary states emerge due to the adaptive feature of the network and classify several bifurcation scenarios in which these states are created and stabilized.
} 
\maketitle

\section{Introduction}

\label{intro} Adaptive networks appear in many real-world applications.
One of the main motivations for studying such networks comes from
the field of neuroscience where the weights of the synaptic coupling
can adapt depending on the activity of the neurons that are involved
in the coupling~\cite{GER96,CLO10}. For instance, the coupling
weights can change in response to the relative timings of neuronal
spiking as it is the case for spike-timing dependent plasticity \cite{MAR97a,ABB00,CAP08a,POP13,LUE16}.
Apart from neuroscience, there are many examples of adaptive networks
in chemical~\cite{JAI01}, social \cite{GRO08a}, and other systems. 

In this paper we consider a model combining both adaptivity and a
complex coupling structure. We restrict our analysis to one specific
coupling structure, a nonlocally coupled ring, on which the coupling weights
are adapted depending on the dynamics of the network. The nonlocal
ring networks, where each node is coupled to all nodes within a certain coupling range, are known to be important systems appearing in many
applied problems and theoretical studies \cite{PAS95,BRE97,YAN08a,BON09,ZOU09b,HOR09b,PER10c,OME11,KAN13,OME13,YAN15a,KLI17,OME18a}. 
For instance, they are important motifs in neural
networks \cite{COM03,SPO11,POP11}.

More specifically, we study rings of nonlocally coupled oscillators
based on the Kuramoto-Sakaguchi model~\cite{SAK86,PIK15}
with an additional adaptation dynamics of the coupling weights. For an all-to-all coupling topology similar models
have been recently studied~\cite{REN07,AOK09,AOK11,PIC11a,TIM14,NEK16,KAS17,AVA18,BER19,BER19a}, but very little is known about the dynamics of these systems if the base topology is more complex~\cite{KAS18a}. On globally coupled networks adaptive Kuramoto-Sakaguchi type models have been shown to exhibit diverse complex dynamical behavior. In
particular, stable multi-frequency clusters emerge in this system,
when the oscillators split into groups of strongly coupled oscillators
with the same average frequency. Such a phenomenon does not occur
in the classical Kuramoto or Kuramoto-Sakaguchi system. The clusters
are shown to possess a hierarchical structure, i.e., their sizes are
significantly different~\cite{KAS17,BER19}. Such a structure leads
to significantly different frequencies of the clusters and, as a result,
to their uncoupling. This phenomenon is also reported for adaptive
networks of Morris-Lecar, Hindmarsh-Rose, and Hodgkin-Huxley neurons with either spike-timing-dependent
plasticity or Hebbian learning rule~\cite{POP15,CHA17a,ROE19a}.
The role of hierarchy and modularity in brain networks has recently
been discussed for real brain networks, as well \cite{BAS08,MEU10a,BAS10,LOH14,BET17,ASH19}.

A particular case of hierarchical multicluster states are solitary
states for which only one single element behaves differently compared
with the behavior of the background group, i.e., the neighboring elements. These states have been
found in diverse dynamical systems such as generalized Kuramoto-Sakaguchi
models~\cite{MAI14a,WU18a,TEI19,CHE19b}, the Kuramoto model with
inertia~\cite{JAR15,JAR18,TAH19}, the Stuart-Landau model~\cite{SAT19}, the
FitzHugh-Nagumo model~\cite{RYB19a,SCH19a}, systems of excitable units~\cite{ZAK16b}
as well as Lozi maps~\cite{RYB17} and even experimental setups
of coupled pendula~\cite{KAP14}. Solitary states are considered
as important states in the transition from coherent to incoherent
dynamics~\cite{JAR15,SEM15b,MIK18}. Despite their appearance in many well-studied
models, the mechanisms of their emergence are less understood. Until
now, only a few works could shed some light on the details 
behind their formation~\cite{MAI14a,JAR18,SEM18a,TEI19}.

In this article, we analyze the emergence of solitary states in the
presence of plastic coupling weights and unveil the bifurcation scenarios
in which solitary states are formed and (de)stabilized. For this,
in Sec.~\ref{sec:model}, we introduce the model and coherence measures
which will be used throughout the article. Numerical results and a
rigorous definition for multi-frequency cluster and solitary states
on complex networks are presented in Sec.~\ref{sec:numerics}. In
Sec.~\ref{sec:onecluster} we provide a more detailed analysis of
one-cluster states. Here, relations between local and global properties
are derived. The salient role of rotating-wave clusters are underlined
and the crucial dependence of their stability on the coupling range
and the wavenumber are rigorously described. Some of the proofs are
presented in the App.~\ref{app:Local2Global} and~\ref{app:StabOneCluster}.
After this, we focus on the analysis of solitary states in Sec.~\ref{sec:solitaries}.
A reduced model for two-clusters is derived and a variety of bifurcations
in which solitary states are born and stabilized are presented. Section~\ref{sec:conclusion}
summarizes our findings.

\section{Model}

\label{sec:model}

We study the system of $N$ adaptively coupled identical phase oscillators
\cite{AOK09,AOK11,NEK16,KAS17,BER19,BER19a} 
\begin{align}
\frac{\mathrm{d}\phi_{i}}{\mathrm{d}t} & =1-\frac{1}{\sum_{j=1}^{N}a_{ij}}\sum_{j=1}^{N}a_{ij}\kappa_{ij}\sin(\phi_{i}-\phi_{j}+\alpha)\label{eqn:AKS_Mdl_phi}\\
\frac{\mathrm{d}\kappa_{ij}}{\mathrm{d}t} & =-\epsilon\left(\kappa_{ij}+\sin(\phi_{i}-\phi_{j}+\beta)\right)\label{eqn:AKS_Mdl_kappa}
\end{align}
where $i,j=1,\dots,N$, $\phi_{i}\in S^{1}$ are the phases of the
oscillators, $a_{ij}\in\{0,1\}$ are the entries of the adjacency
matrix $A$ determining the base topology, $\kappa_{ij}\in[-1,1]$
are slowly changing adaptive coupling strengths, $0<\epsilon\ll1$
is the rate of the adaptation, and $\alpha$, $\beta$ are coupling
and adaptation phase lags. Note that the natural oscillation frequency has been normalized to $1$ by the rotating coordinate frame. 

The base topology $A$ determines the structure of the network, on
which the adaptation takes place. Equation (\ref{eqn:AKS_Mdl_kappa})
for the adaptation is used only for ``active'' weights $\kappa_{ij}$
corresponding to $a_{ij}=1$. Similarly, the sum in (\ref{eqn:AKS_Mdl_phi})
goes over these links. Here we consider the topology of a nonlocally coupled ring
given by 
\begin{equation}
a_{ij}=\begin{cases}
1\quad\mbox{for}\ \ 0<(i-j)\,\mbox{mod}\,N\le P,\\
0\quad\mbox{otherwise.}
\end{cases}\label{eq:ring}
\end{equation}
This means that any two oscillators are coupled if their indices $i$
and $j$ are separated at most by the coupling radius $P$. The coupling Eq.~(\ref{eq:ring}) defines a nonlocal ring structure with coupling range $P$ to each side and two special
limiting cases: local ring for $P=1$ and globally coupled network
for $P=N/2$ (if $N$ is even, else $P=(N+1)/2$). The matrix of the
form (\ref{eq:ring}) is circulant~\cite{GRA06} and has constant row sum, i.e., $\sum_{j=1}^{N}a_{ij}=2P$ for all $i=1,\dots,N$.

Let us mention important properties of the model~\eqref{eqn:AKS_Mdl_phi}--\eqref{eqn:AKS_Mdl_kappa}.
The small parameter $\epsilon$ separates the time scales of the slowly
adapting coupling weights from the fast moving phase oscillators.
Further, the coupling weights are confined to the interval $-1\le\kappa_{ij}\le1$
due to the fact that $\mathrm{d}\kappa_{ij}/\mathrm{d}t\le0$ for
$\kappa_{ij}=1$ and $\mathrm{d}\kappa_{ij}/\mathrm{d}t\ge0$ for
$\kappa_{ij}=-1$, see Ref.~\cite{KAS17}. Due to the invariance
of system~\eqref{eqn:AKS_Mdl_phi}--\eqref{eqn:AKS_Mdl_kappa} with
respect to the shift $\phi_{i}\mapsto\phi_{i}+\psi$ for all $i=1,\dots,N$
and $\psi\in[0,2\pi)$, the natural frequency of each oscillator can
be set to any value, which we choose to be zero. Finally due to the
following symmetries with respect to parameters $\alpha$ and $\beta$:
\begin{align*}
(\alpha,\beta,\phi_{i},\kappa_{ij}) & \mapsto(-\alpha,\pi-\beta,-\phi_{i},\kappa_{ij}),\\
(\alpha,\beta,\phi_{i},\kappa_{ij}) & \mapsto(\alpha+\pi,\beta+\pi,\phi_{i},-\kappa_{ij}),
\end{align*}
the analysis can be restricted to the parameter regions $\alpha\in[0,\pi/2)$
and $\beta\in[-\pi,\pi)$.

Two measures of coherence will be used in the paper: first, the $n$th
moment of the $i$th ($i=1,\dots,N$) complex local order parameter
as given by 
\begin{align}
Z_{i}^{(n)}(\bm{\phi}):=\frac{1}{2P}\sum_{j=1}^{N}a_{ij}e^{\mathrm{i}n\phi_{j}}=R_{i}^{(n)}(\bm{\phi})e^{\mathrm{i}\vartheta_{i}^{(n)}(\bm{\phi})}\label{eqn:local_order}
\end{align}
where $\bm{\phi}=(\phi_{1},\dots,\phi_{N})^{T}$ , $R_{i}^{(n)}(\bm{\phi})$
is the $n$th local order parameter and $\vartheta_{i}^{(n)}(\bm{\phi})$
the $n$th local mean-phase; second, the complex (global) order parameter
\begin{align}
Z^{(n)}(\bm{\phi}):=\frac{1}{N}\sum_{j=1}^{N}e^{\mathrm{i}n\phi_{j}}=R^{(n)}(\bm{\phi})e^{\mathrm{i}\vartheta^{(n)}(\bm{\phi})}\label{eqn:global_order}
\end{align}
where $R^{(n)}(\bm{\phi})$ is the $n$th (global) order parameter
and $\vartheta^{(n)}(\bm{\phi})$ the $n$th (global) mean-phase.
Both measures are used throughout the article to characterize asymptotic
states of \eqref{eqn:AKS_Mdl_phi}--\eqref{eqn:AKS_Mdl_kappa}. 


\section{Numerical observations: one-cluster, multiclusters, and solitary
states}

\label{sec:numerics}

This section is devoted to the numerical analysis of system~\eqref{eqn:AKS_Mdl_phi}--\eqref{eqn:AKS_Mdl_kappa}
and the description of several dynamical states which occur for this
system. More specifically, we report one-cluster, multi-cluster, and
solitary states. While this section describes the states in a phenomenological
fashion, more rigorous results are presented in the subsequent sections~\ref{sec:onecluster}--\ref{sec:solitaries}.

Note that the observed one-cluster and multicluster states are similar
to those reported in \cite{BER19,BER19a} for the all-to-all coupling base
topology. However, there are important differences due to the ring structure of our system, which will be discussed in detail. 

For the numerical simulations in this section a system of $N=100$
oscillators with coupling range $P=20$ is studied. The value of $\epsilon$
is set to $0.01$ and the system parameter $\alpha$ and $\beta$
are varied in the ranges $[0,\pi/2]$ and $[-\pi,\pi]$, respectively.
All results are obtained starting from uniformly distributed random
initial conditions and a simulation time of $t=20000$. Several types
of synchronization patterns are found in the numerical simulations,
depending on the values of $\alpha$ and $\beta$. 

\begin{figure}
\centering \includegraphics{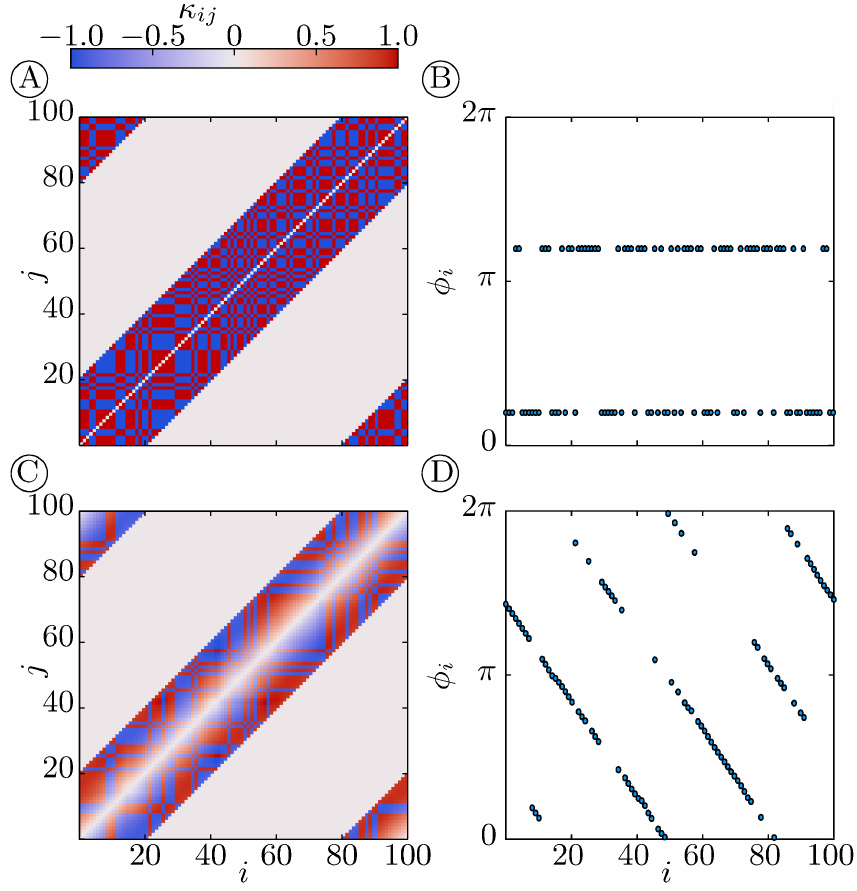} \caption{\label{fig:1cl} Illustration for two types of one-cluster states.
The panels (A,C) show the asymptotic coupling matrices and (B,D) snapshots of
the phases at a fixed time. Results for the one-cluster states of
antipodal type are presented in (A,B) where $\alpha=0.19\pi$, $\beta=-0.66\pi$
and of splay type in (C,D) where $\alpha=0.35\pi$, $\beta=0.01\pi$. Parameters: $N=100$, $P=20$, $\epsilon=0.01$.}
\end{figure}

\subsection{One-cluster states}

A one-cluster is defined as a frequency synchronized state with 
$$\phi_{i}=\Omega t+\chi_{i},\quad i=1,\dots,N $$
with a collective frequency $\Omega\in\mathbb{R}$
and individual phase shifts $\chi_{i}\in[0,2\pi)$ \cite{BER19,BER19a}.
The two types of one-cluster states found in the numerical simulations
are either of \emph{antipodal} or \emph{splay type} whose asymptotic
configurations are displayed in Fig.~\ref{fig:1cl}(A,B) and Fig.~\ref{fig:1cl}(C,D),
respectively. The antipodal and splay-type clusters have been introduced
previously in \cite{BER19}. In the antipodal cluster, all phases
$\phi_{i}$ are either in-phase or in anti-phase, i.e., $\chi_{i}\in\{\chi,\chi+\pi\}$
with $\chi\in[0,2\pi)$ and hence $R^{(2)}(\bm{\phi})=1$. In the
splay cluster the phases are distributed across the interval $[0,2\pi)$
such that the global second order parameter, as defined in equation~\eqref{eqn:global_order},
vanishes, i.e., $R^{(2)}(\bm{\phi})=0$. In Fig.~\ref{fig:1cl}(A,C) the
coupling structures corresponding to the two types of one-clusters
are displayed. Note that the coupling weights are solely described
by the phase differences of the oscillators and are given by  \[\kappa_{ij}=-\sin(\chi_{i}-\chi_{j}+\beta).\] The one-cluster
states, which exist in our ring case and the all-to-all base topology
case from \cite{BER19,BER19a} have the same representation except
for the fact that some of the coupling weights are absent in the case
of the ring, see empty entries in Fig.~\ref{fig:1cl}(A,C). 

\subsection{Multicluster states}

As described in~\cite{BER19,BER19a}, one-clusters can serve as building
blocks for multi-frequency clustered states where the phase dynamics and
the coupling matrix $\kappa$ are divided into different groups; 
$\kappa_{ij,\mu\nu}$ 
refers to the coupling weight for the connection from the $i$th oscillator of
the $\mu$th cluster to the $j$th oscillator of the $\nu$th cluster.
Analogously, $\phi_{i,\mu}$ denotes the $i$th phase oscillator in
the $\mu$th cluster. The temporal behavior for each oscillator in an
$M$-cluster state takes the form 
\begin{align}
\phi_{i,\mu}(t)=\Omega_{\mu}t+\chi_{i,\mu}+s_{i,\mu}(t) &  & \begin{split}\mu=1,\dots,M\\
i=1,\dots,N_{\mu}
\end{split}
\label{eq:MCstate_gen}
\end{align}
where $M$ is the number of clusters, $N_{\mu}$ is the number of
oscillators in the $\mu$th cluster, $\chi_{i,\mu}\in[0,2\pi)$ are
phase lags, and $\Omega_{\mu}\in\mathbb{R}$ is the collective frequency
of the oscillators in the $\mu$th cluster. The functions $s_{i,\mu}(t)$
are assumed to be bounded.

Both types of one-clusters give rise to multi-frequency
cluster states, see Fig.~\ref{fig:1cl}, similarly to the case of all-to-all base coupling. However, in case of more complex network structures
the definition of a frequency-cluster has to be refined to account
for the connectedness of the individual building blocks. Therefore,
\emph{a multi-frequency-cluster (shortly: multicluster) consists of groups
of frequency synchronized oscillators for which the subnetwork (or
subgraph), induced by the individual groups of nodes, is connected}.
Here, we say a network is connected if there is directed path from
each node to every other node of the network. In case of a directed
graph, we require the property of weak connectedness for the induced
subgraph for the cluster. For an introduction to the terminology we
refer the reader to~\cite{KOR18}. 

\begin{figure}
\centering \includegraphics{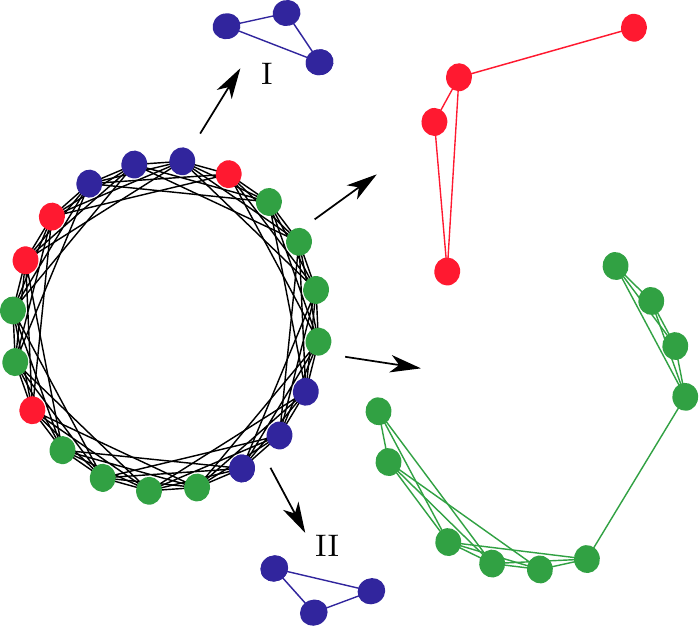} \caption{Schematic figure illustrating the definition of multicluster and subnetworks
induced by groups of nodes with the same average frequency. The full
network (left) consists of $N=20$ nodes and has a nonlocal ring structure
with $P=4$. The colors of the nodes indicate their average frequencies.
Clusters are shown by the equally colored nodes that form connected
sub-networks. Even though the two blue groups I and II possess the
same averaged frequencies, they form two different clusters, since
they are not connected. \label{fig:subnetworks} }
\end{figure}

Let us illustrate the above definition. Consider a nonlocal ring network
of $N=20$ oscillators with coupling range $P=4$ as presented in
Fig.~\ref{fig:subnetworks}. Suppose that for each node of the network
we have a certain average frequency which is indicated by the color.
In Fig.~\ref{fig:subnetworks}, we have three different average frequencies
denoted by the green, blue, and red colors. The individual clusters
are given by the connected subnetworks induced by equally colored
nodes. Note that even though the blue nodes have the same average
frequency, they are forming two different clusters (\rom{1},\rom{2})
corresponding to two connected components. Note that the induced subnetworks
are not necessarily regular even if the base topology is regular,
see for instance the red or green subnetworks.

\begin{figure}
\centering \includegraphics{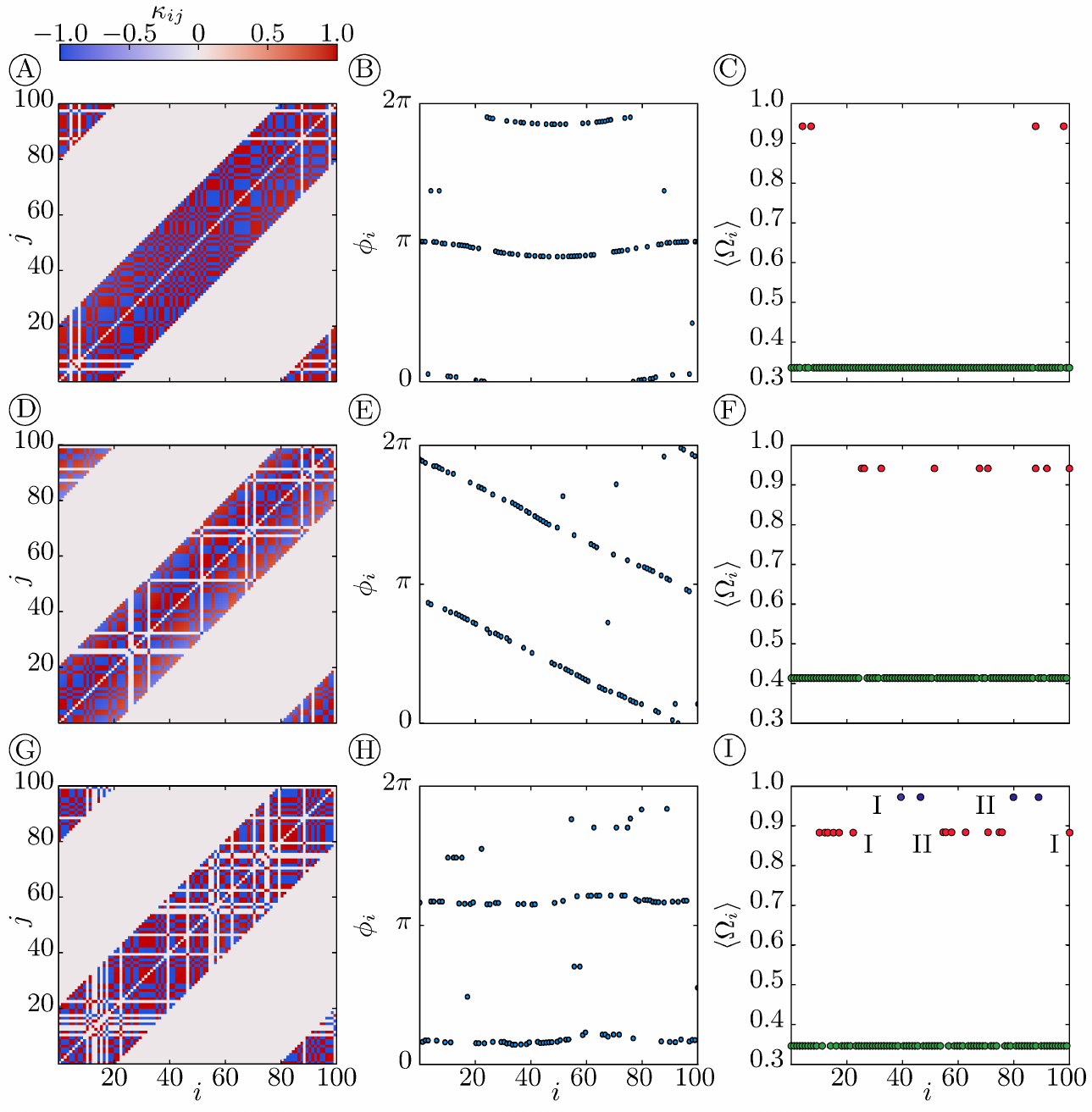} \caption{\label{fig:multicl} Illustration of the different types of multicluster
states. The panels (A,D,G) show the coupling matrix, (B,E,H) phase
snapshots and (C,F,I) average frequencies. (A-C): antipodal two-cluster
for $\alpha=0.23\pi$, $\beta=-0.56\pi$; (D-F): splay two-cluster
for $\alpha=0.19\pi$, $\beta=-0.45\pi$; (G-I): antipodal five-cluster (\rom{1}, \rom{2} denote the two connected components of the red and the blue clusters)
for $\alpha=0.3\pi$, $\beta=-0.53\pi$. Parameters: $N=100$, $P=20$, $\epsilon=0.01$.}
\end{figure}

Figure~\ref{fig:multicl} shows two-cluster states of antipodal (Figs.~\ref{fig:multicl}(A--C))
and splay type (Figs.~\ref{fig:multicl}(D--F)). In both cases,
Figs.~\ref{fig:multicl}(C,F) show that there are two distinct groups
of oscillators with different averaged frequencies (green, red). It
is easily verified that these groups form connected subnetworks and,
hence, they form a two-cluster state. Due to the frequency difference
between the individual clusters, the groups of oscillators decouple
effectively. This can be seen in Figs.~\ref{fig:multicl}(A,D) where
only oscillators of the same cluster are strongly coupled compared
to the coupling between the clusters given by the respective coupling
weights $\kappa_{ij,\mu\mu}$ and $\kappa_{ij,\mu\nu}\approx0$ ($\mu\ne\nu$).

Snapshots of the phase distributions are presented in~Figs~\ref{fig:multicl}(B,E).
Figure~\ref{fig:multicl}(B) is showing the phase distribution of
an antipodal cluster. In contrast to the case of global coupling~\cite{BER19},
the phases do not posses the exact antipodal property anymore. In
fact, the scattering of the antipodal phase distribution is caused
by the structure of the induced subnetwork which is not regular anymore
for the in-degree of each node, i.e., the subnetwork does not have
constant row sum, see e.g. Fig.~\ref{fig:subnetworks}. Note that
in case of a global base topology, all induced subnetworks are global.
The phase snapshot for the splay multicluster is displayed in Fig.~\ref{fig:multicl}(E).
Here, as in the case with global base structure, the phase distribution
posses the property that $R^{(2)}(\bm{\phi})=0$.

Another, more complex, antipodal five-cluster is presented in Fig.~\ref{fig:multicl}(G--I).
In Fig.~\ref{fig:multicl}(I), we observe three groups, a big one
(green) and two smaller ones (red, blue), with different frequencies.
Moreover, in accordance with the definition of a frequency cluster
and the illustration in Fig.~\ref{fig:subnetworks}, the blue and
the red groups possesses two connected components (\rom{1},\rom{2})
each. This fact implies the presence of five individual frequency
clusters in Fig.~\ref{fig:multicl}(G--I). Remarkably, the red clusters
I and II as well as the blue clusters I and II are of the same size.
This observation is in contrast to the hierarchical structures discovered
and analyzed in~\cite{KAS17,BER19a}. Hence, in the case of the ring
base structure, the evenly sized clusters can appear, which was not
possible in the case of global base structure \cite{BER19,BER19a,KAS17}
and identical oscillators. 

\begin{figure}
\centering \includegraphics{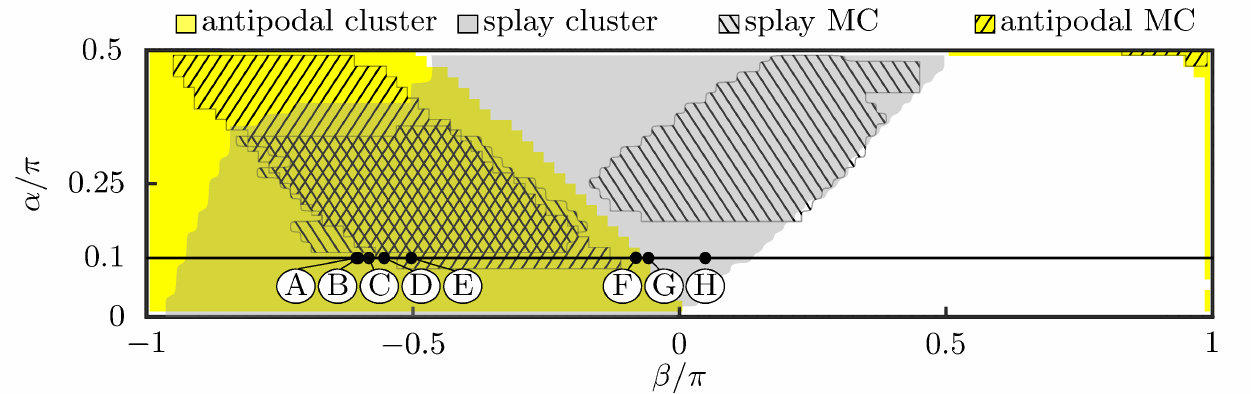} \caption{Map of regimes for one- and multicluster (MC) states of antipodal and
splay type in $(\alpha,\beta)$ parameter space. Parameters: $N=100$, $P=20$, $\epsilon=0.01$. The horizontal black line at $\alpha=0.1$
shows the location for the parameter $\beta$ where the emergence
of solitary states is analyzed, see Fig.~\ref{fig:emergence_solitaries}
in Sec.~\ref{sec:solitaries}}
\label{fig:map_regimes} 
\end{figure}

While several examples for one- and multicluster states have been
described above, Fig.~\ref{fig:map_regimes} shows that these
states are observable in a wide range in the $(\alpha,\beta)$ parameter
space. The diagram in Fig.~\ref{fig:map_regimes} is produced by
running simulations of~\eqref{eqn:AKS_Mdl_phi}--\eqref{eqn:AKS_Mdl_kappa}
from random initial conditions. In case a one-cluster or multicluster
is found, the region is colored or hatched, respectively, in accordance
with the legend in Fig.~\ref{fig:map_regimes}. We further used a
continuation method in the $(\alpha,\beta)$ parameter space to show
the full extent where the various types of multiclusters can be observed.
In section~\ref{sec:equlibria} and~\ref{sec:StabilityAnalysis},
we provide a more rigorous description for the existence and stability
properties of the one-clusters.

\begin{figure}
\centering \centering \includegraphics{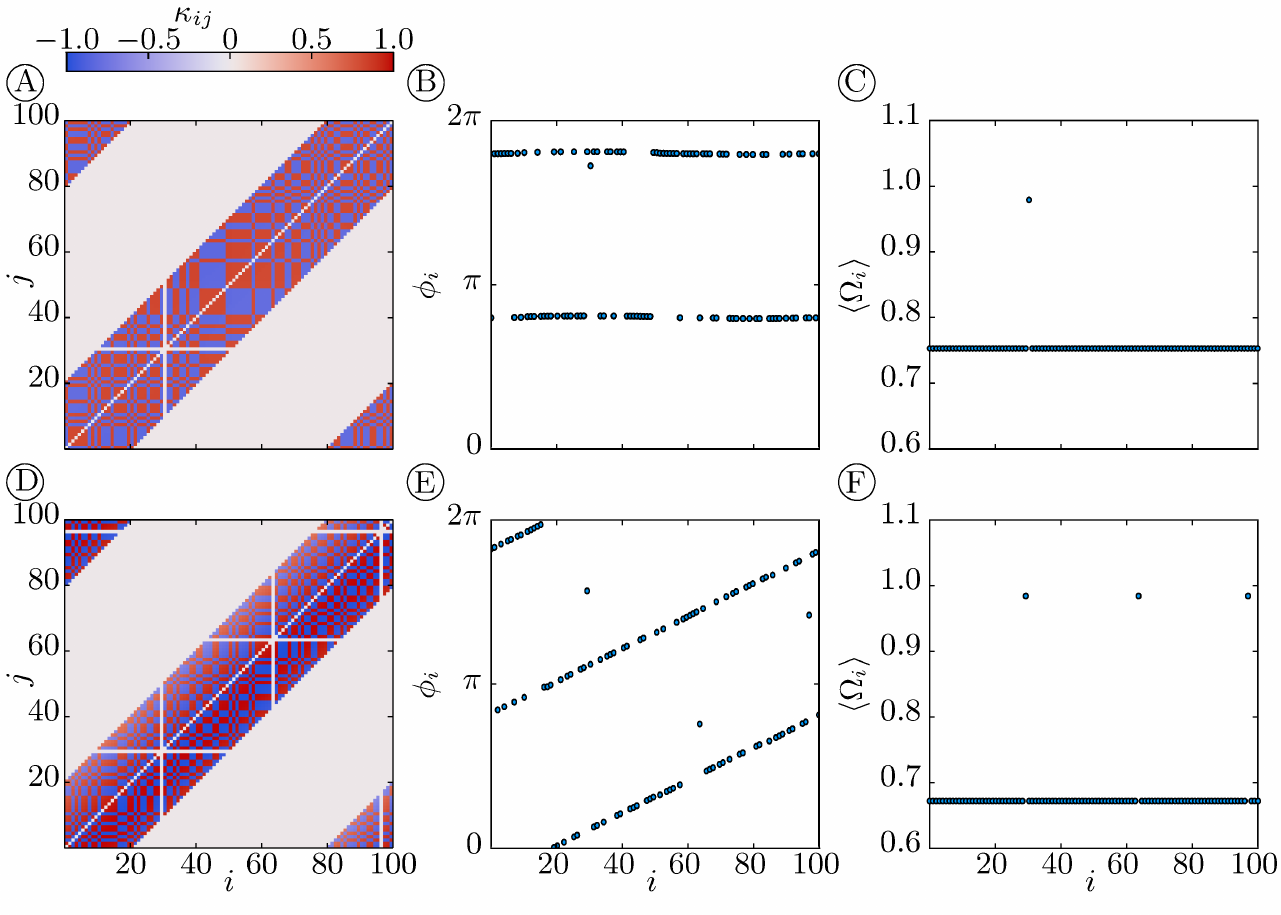} \caption{\label{fig:solitaries} Illustration of solitary states. The panels
(A,D) show coupling matrix, (B,E) phase snapshots, and (C,F) average
frequencies. (A-C): single solitary state for $\alpha=0.1\pi$, $\beta=-0.3\pi$;
(D-F): three uncoupled solitary states for $\alpha=0.15\pi$, $\beta=-0.41\pi$. Parameters: $N=100$, $P=20$, $\epsilon=0.01$.}
\end{figure}

\subsection{Solitary states}

For systems with global base coupling, the clusters in the multicluster
states were found to be hierarchical in nature, i.e. the clusters
varied in size significantly \cite{KAS17,BER19,BER19a}. As we have
mentioned above, in the nonlocal base coupling case, multicluster
states with one large and many smaller, similar in size, clusters
have been observed. Figure~\ref{fig:solitaries} shows a particular
example of this phenomenon, called solitary states, where either one
single oscillator (upper panels) or three single oscillators (lower
panels) decouple from a large cluster. The solitary states are particular
examples of multiclusters with a large group of frequency synchronized
oscillators (background cluster) and individual solitary nodes with
different frequency, i.e., clusters consisting of only one oscillator.
These special kind of states, for which we provide an analysis of
their emergence in Sec.~\ref{sec:solitaries}, are of particular
interest as they are found in various dynamical systems~\cite{MAI14a,ASH15,SEM15b,WOJ16,PRE16,MAI17,JAR18,TEI19,TAH19}.

\section{Analysis of one-cluster states}
\label{sec:onecluster}

\subsection{Existence and classification of one-cluster states}
\label{sec:equlibria}

In this section we study the antipodal and splay one-cluster states
in more details. Due to the $S^{1}$ symmetry of system (\ref{eqn:AKS_Mdl_phi})-(\ref{eqn:AKS_Mdl_kappa}),
the following phase-locked solutions appear generically 
\begin{align}
\phi_{i}(t)=\Omega t+\chi_{i}, & i=1,\dots,N,\label{eq:1C_phi}
\end{align}
where $\chi_{i}\in[0,2\pi)$ are fixed phase lags and $\Omega$ the
cluster frequency. It is clear that such solutions describe a one-cluster
state since the frequencies are the same. By substituting \eqref{eq:1C_phi}
into \eqref{eqn:AKS_Mdl_phi}--\eqref{eqn:AKS_Mdl_kappa}, we obtain
\begin{align}
\kappa_{ij}=-\sin(\chi_{i}-\chi_{j}+\beta),\label{eq:1C_kappa}
\end{align}
\begin{equation}
\Omega=\frac{1}{2}\cos(\alpha-\beta)-\frac{1}{4P}\sum_{j=i-P}^{i+P}\cos(2\chi_{i}-2\chi_{j}+\alpha+\beta).\label{eq:Omega}
\end{equation}
The equation (\ref{eq:Omega}) implies that the one-cluster state
exists only if the following expression is independent of the index
$i$ 
\begin{align}
\frac{1}{2P}\sum_{j=i-P}^{i+P}\cos(2\chi_{i}-2\chi_{j}+\alpha+\beta)
=\Re\left(R_{i}^{(2)}e^{\mathrm{i}(\vartheta_{i}^{(2)}-2\chi_{i}-\alpha-\beta)}\right).\label{eq:const}
\end{align}
Equation (\ref{eq:const}) allows for the distinction of two types of distributions
of the phase-lags for which it is independent of $i$. We call a cluster
of 
\begin{description}
\item [{\emph{(i)}}] \emph{Antipodal type}, if $\chi_{i}\in\{0,\pi\}$
for all $i=1,\dots,N$. In this case, the sum in (\ref{eq:const})
equals $\cos(\alpha+\beta)$, and of
\item [{\emph{(ii)}}] \emph{Local splay type}, if $\chi_{i}\in[0,2\pi)$
are such that the second order parameter $Z_{i}^{(2)}=R_{i}^{(2)}(\bm{\chi})e^{\mathrm{i}\vartheta_{i}^{(2)}}$
satisfies 
\begin{align}
R_{i}^{(2)}(\bm{\chi}) & =R_{c}^{(2)}(\bm{\chi})\label{eq:lss}\\
\vartheta_{i}^{(2)} & =2\chi_{i}+\chi_0\label{eq:lss2}
\end{align}
with $0\le R_{c}^{(2)}(\bm{\chi})<1$ and $\chi_0\in[0,2\pi)$ independent
on $i$.
\end{description}
Note that the additional degree of freedom for clusters of local splay type, i.e. $\chi_0\in[0,2\pi)$ arbitrary, is due to the $S^1$ symmetry. The constant $\chi_0$ can be set to $0$ without loss of generality.
The both types of phase distribution lead to one-cluster states for
the system~\eqref{eqn:AKS_Mdl_phi}--\eqref{eqn:AKS_Mdl_kappa}
with frequency 
\begin{align}
\Omega=\begin{cases}
\sin\alpha\sin\beta & \,\text{antipodal type},\\
\frac{1}{2}\left(\cos(\alpha-\beta)-R_{c}^{(2)}\cos(\alpha+\beta)\right) & \,\text{local splay type}.
\end{cases}\label{eq:1C_freq_}
\end{align}
In the context of globally coupled base topologies, antipodal and
splay type phase distribution have been extensively discussed~\cite{ASH08,BER19,BER19a},
where the (global) splay clusters, also called fuzzy clusters~\cite{MAI14a},
are defined by the global condition $Z^{(2)}(\boldsymbol{\mathbf{\chi}})=0$.
Remarkably, if a phase distribution is of the local splay type (as
described by (\ref{eq:lss})-(\ref{eq:lss2})) then it is of splay
type as well, see Appendix~\ref{app:Local2Global} for more details.
The converse is not true in general. Hence, the class of local splay
clusters is ''smaller'' than the class of
global splay clusters. In addition, local splay clusters do not necessarily
form families of solutions. According to the definition of local splay
cluster, generically $N$ complex algebraic equations have to be solved
for $N$ unknown phase-lags $\chi_{i}$. Therefore, the set of equations
for the phase-lags is overdetermined and the set of local splay states
might be empty. However, it is not the case due to the symmetries
of the system and the base coupling structure. The symmetry of the nonlocal ring structure allows for
constructing explicit, symmetric examples for the clusters of local
splay type. These are clusters of the rotating-wave type. 

\setcounter{enumi}{2} 
\begin{description}
\item [{\emph{(ii')}}] The clusters are of\emph{ rotating-wave type}, if
$\chi_{i}=ik\frac{2\pi}{N}$, where $k=1,\dots,N$ is the wavenumber. In the literature, the notion "splay state" is often restricted to this definition. 
\end{description}
Let us show that the rotating wave clusters \emph{(ii')} are the local
splay states \emph{(ii)}. For this we write the phase distribution
as $\bm{\chi}_{k}=(2\pi k/N,\dots2\pi k(N-1)/N,0)^{T}$. Then, we
have 
\begin{align}
Z_{i}^{(n)}(\bm{\chi}_{k})=\frac{1}{2P}\sum_{j=i-P}^{i+P}e^{\mathrm{i}nkj\frac{2\pi}{N}}=e^{\mathrm{i}nki\frac{2\pi}{N}}R_{N}^{(n)}(\bm{\chi}_{k}) & ,
\end{align}
where 
\begin{align}
R_{N}^{(n)}(\bm{\chi}_{k})=\frac{1}{P}\left(\sum_{j=1}^{P}\cos(nkj\frac{2\pi}{N})\right).
\end{align}
we conclude that all rotating-wave states with $k\ne0,N/2$ are local
splay states and thus solutions to~\eqref{eqn:AKS_Mdl_phi}--\eqref{eqn:AKS_Mdl_kappa}.
Rotating-wave clusters with $k=0,N/2$ are of antipodal type. The
$n$th moment local order parameter $\left|Z_{i}^{(n)}(\bm{\chi}_{k})\right|=R_{N}^{(n)}(\bm{\chi}_{k})$
is constant for all $i=1,\dots,N$ and its value depends on the wavenumber
$k$. 

Note further that $Z_{i}^{(n)}(\bm{\chi}_{k})=Z_{i}^{(nk)}(\bm{\chi}_{1})$,
which connects the moment of the order parameters with the wavenumber
of the rotating-wave states. For globally coupled base structures,
the rotating-wave states are found to be very important in describing
the main features of antipodal and global splay type clusters such
as stability. The next section is devoted to the description of the
stability condition for rotating-wave states.

\subsection{Stability of one-cluster states\label{sec:StabilityAnalysis}}

In the following, the stability of one-clusters is analyzed. In order
to study the local stability of one-cluster solutions described in
Sec.~\ref{sec:equlibria}, we linearize the system of differential
equations \eqref{eqn:AKS_Mdl_phi}--\eqref{eqn:AKS_Mdl_kappa} around
the phase-locked states 
\begin{align*}
\phi_{i}(t) & =\Omega t+a_{i},\\
\kappa_{ij} & =-\sin(a_{i}-a_{j}+\beta).
\end{align*}
These solutions are equilibria relative to the $S^{1}$ symmetry \cite{GOL88a},
therefore the linearization around such solutions leads to a linear
system with constant coefficients, despite the time dependency of
$\phi(t)$. Practically, one can first move to the co-rotating coordinate
system by introducing the new variable $\phi(t)-\Omega t$ and then
linearize around the equilibrium in the new coordinates. As a result,
we obtain the following linearized system for the perturbations $\delta\phi_{i}$
and $\delta\kappa_{ij}$ 
\begin{multline}
\frac{d}{dt}\delta\phi_{i}=\frac{1}{4P}\sum_{j=i-P}^{i+P}\left[\sin(\beta-\alpha)+\sin(2(a_{i}-a_{j})+\alpha+\beta)\right]\left(\delta\phi_{i}-\delta\phi_{j}\right)\\
-\frac{1}{2P}\sum_{j=i-P}^{i+P}\sin(a_{i}-a_{j}+\alpha)\delta\kappa_{ij},\label{eq:Linearized_phi}
\end{multline}
\begin{align}
\frac{d}{dt}\delta\kappa_{ij}=-\epsilon a_{ij}\left(\delta\kappa_{ij}+\cos(a_{i}-a_{j}+\beta)\left(\delta\phi_{i}-\delta\phi_{j}\right)\right).\label{eq:Linearized_kappa}
\end{align}

System (\ref{eq:Linearized_phi})-(\ref{eq:Linearized_kappa}) is
a $(N+N^{2})$-dimensional linear system of ordinary differential equations,
which can be written in the form $\boldsymbol{x}'=\boldsymbol{L}\boldsymbol{x}$,
$\boldsymbol{x}\in\mathbb{R}^{N+N^{2}}$, and the stability of which
is determined by the eigenvalues of the matrix $\boldsymbol{L}$.
For all antipodal and rotating-wave states the stability analysis
can be done explicitly. However, the calculations are quite lengthy (see Appendix~\ref{app:StabOneCluster}). Summarizing
the results of these calculations, the spectrum $S$ of the eigenvalues,
corresponding to the rotating-wave one-clusters, is given by 
\begin{align}
S=\left\{ 0,-\epsilon,\left\{ \lambda_{l,1}\right\} _{l=1}^{N},\left\{ \lambda_{l,2}\right\} _{l=1}^{N}\right\} .\label{eq:RotWave_LyaSpectrum}
\end{align}
Here $\lambda_{l,1}$ and $\lambda_{l,2}$ are the solutions of the
quadratic equation 
\begin{align}
\lambda_{l}^{2}-\frac{\lambda_{l}}{2}\left[L(\alpha,\beta,l,k)+(R_{N}^{(l)}(\boldsymbol{\chi}_{1})-1)\sin(\alpha-\beta)-2\epsilon\right]-\epsilon L(\alpha,\beta,l,k)=0,\label{eq:quadratic_lyap}
\end{align}
where the complex function $L$ as defined in Eq.~\eqref{eq:def_Lfunction}
in Appendix~\ref{app:StabOneCluster}, $k$ is the wavenumber. 

\begin{figure}
\centering \includegraphics{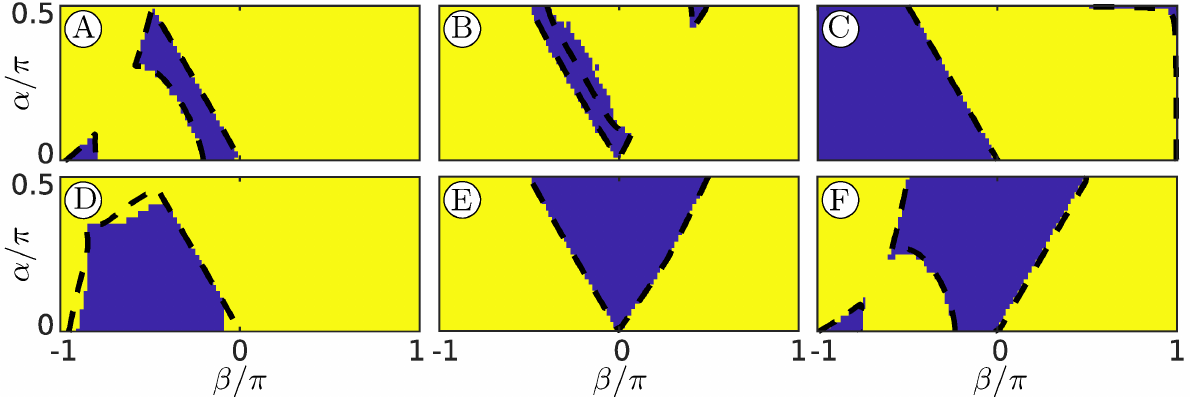} \caption{\label{fig:1cl_stab} Stability of one-cluster states for different
wavenumbers $k$ and coupling ranges $P$. Regions of stability for
the one-cluster states are colored in blue, while instability in yellow.
The borders of stability (black dashed lines) are obtained from the
eigenvalues \eqref{eq:RotWave_LyaSpectrum}. Parameters are as follows:
(A) $P=10$, $k=1$; (B) $P=10$, $k=4$; (C) $P=10$, $k=25$; (D)
$P=5$, $k=1$; (E) $P=20$, $k=1$; and (F) $P=25$, $k=1$. The
other parameters are $N=50$ and $\epsilon=0.01$.}
\end{figure}

In Figure~\ref{fig:1cl_stab} we show the stability of rotating-wave
one-clusters in an ensemble of $N=50$ oscillators. The figures demonstrate
regions of stability in the $(\alpha,\beta)$ parameter plane for
different wavenumber $k$ and coupling range $P$. The stability is
obtained by numerical simulations as well as by the Lyapunov spectrum Eq.~\eqref{eq:RotWave_LyaSpectrum}.
The borders of stability, as they are provided by the analytical results,
are displayed with a dashed black line. Numerically the stability
was computed as follows: (i) the theoretical shape of the one-cluster
given by~\eqref{eq:1C_phi}--\eqref{eq:1C_kappa} is used as initial
conditions with a small random perturbation in the range of $[-0.01,0.01]$;
(ii) then we solve the system numerically for $t=20000$ time units;
(iii) compute the euclidean norm between the initially perturbed state
and the theoretical one as well as between the final state after $t=20000$
and the theoretical one; (iv) in case the second norm is smaller than
the first, meaning that the trajectory approaches the theoretical
one-cluster state, we consider the one-cluster state as stable and
color the corresponding region in blue, otherwise the state is considered
as unstable and the corresponding region is colored in yellow.

The diagrams in the first row of Fig.~\ref{fig:1cl_stab} show
the influence of the wavenumber on the stability of one-clusters.
Here, the coupling range is fixed to $P=10$. For adaptively coupled
phase oscillators with a global base topology, it has been shown that
the stability of rotating-waves of local splay type, i.e., $k\ne0,N/2$,
does not depend on the wavenumber~\cite{BER19,BER19a}. However.
we observe that in case of a nonlocal base structure the shape of
the stability regions crucially depends on the wavenumber $k$, see
Fig.~\ref{fig:1cl_stab}(A,B,C). Moreover, there are no common regions
of stability for the one-cluster states in Fig.~\ref{fig:1cl_stab}(B,C),
i.e., the region of stability in both figures have an empty intersection. 

The diagrams in Fig.~\ref{fig:1cl_stab}(D--F) exemplify the influence
of the coupling range on the stability of the rotating-wave cluster
with $k=1$. We see that in comparison with Fig.~\ref{fig:1cl_stab}(A),
the regions of stability change significantly. Note further that for
$P=25$ the stability regions resemble the results known for the globally
coupled base topology \cite{BER19,BER19a}.

In contrast to the
local splay type clusters, the stability regions for the antipodal
one-cluster states are the same which can be derived from the following. For antipodal states,
the quadratic equation~\eqref{eq:quadratic_lyap} simplifies to 
\begin{align}
\lambda_{l}^{2}-\frac{\lambda_{l}}{2}\left[(1-R_{N}^{(l)}(\boldsymbol{\chi}_1))\sin(\beta)\cos(\alpha)-2\epsilon\right]-\epsilon(1-R_{N}^{(l)}(\boldsymbol{\chi}_1))\sin(\alpha+\beta)=0.\label{eq:quadratic_lyap_antipodal}
\end{align}
since $L(\alpha,\beta,l,k)=\sin(\alpha+\beta)(1-R_{N}^{(l)}(\boldsymbol{\chi}_1))$ for
$k=0,N/2$. 

The regions of stability in Fig.~\ref{fig:1cl_stab}(C)
for a nonlocal coupling structure are in agreement with the regions
of stability found for a global coupling structure~\cite{BER19}.
However, note that Eq.~\eqref{eq:quadratic_lyap_antipodal} differs analytically from the expression found in~\cite{BER19}. The similarity in the stability regions is only due to the small value of $\epsilon$ and the differences would be more pronounced in the presence of larger $\epsilon$. Note further, that in case of $P=N/2$, which is equivalent to global coupling,
the local order parameter $R^{(l)}$ is either $1$ for $l=0$ and
$0$ otherwise. This agrees with the findings in Ref.~\cite{BER19}.

Our stability analysis shows how strongly the ring network structure
and confined coupling range alter the stability properties of the clusters.
Since the analysis in Appendix~\ref{app:StabOneCluster} is not restricted
to nonlocal coupling structures, it provides the analytic tools to
study the influence of more general complex base topologies on the
stability of rotating-wave states.

\section{Emergence of solitary states}

\label{sec:solitaries} In this section, we unveil the mechanism behind
the formation of solitary states as they are illustrated in Fig.~\ref{fig:solitaries}.
As \emph{solitary state} we define the state where all oscillators
in the system are frequency synchronized except one single oscillator,
or several oscillators which do not share their local neighborhood. The majority cluster of the synchronized oscillators
is also called background cluster. 

Let us restrict ourselves to the analysis of a solitary cluster interacting
with an in-phase synchronous cluster (see~\eqref{eq:1C_phi} and~\eqref{eq:1C_kappa}
with $\chi_{i}=0$). Imposing the assumptions, we end-up with the following
$4$-dimensional model where $\phi$ and $\psi$ describe the dynamics
of the solitary cluster and the background in-phase synchronized cluster,
respectively, which are dynamically coupled through $\kappa_{1}$
and $\kappa_{2}$: 
\begin{align*}
\dot{\phi} & =1-\frac{N-1}{N}\kappa_{1}\sin(\phi-\psi+\alpha),\\
\dot{\psi} & =1+\frac{N-2}{N}\sin\alpha\sin\beta-\frac{1}{N}\kappa_{2}\sin(\phi-\psi-\alpha),\\
\dot{\kappa}_{1} & =-\epsilon\left(\kappa_{1}+\sin(\phi-\psi+\beta)\right),\\
\dot{\kappa}_{2} & =-\epsilon\left(\kappa_{2}-\sin(\phi-\psi-\beta)\right).
\end{align*}
The latter equations can be simplified by introducing the phase difference
$\theta=\phi-\psi$ as well as by considering a large ensemble of
oscillators ($N\to\infty$). We obtain the following two-dimensional
system for the dynamics of two clusters, one of which is solitary,
in the large ensemble limit: 
\begin{align}
\dot{\theta} & =-\sin\alpha\sin\beta-\kappa\sin(\theta+\alpha),\label{eq:2Cl_red_system_theta}\\
\dot{\kappa} & =-\epsilon(\kappa+\sin(\theta+\beta)),\label{eq:2Cl_red_system_kappa}
\end{align}
where we denote $\kappa=\kappa_{1}$.

\begin{figure}
\centering \includegraphics[width=1\linewidth]{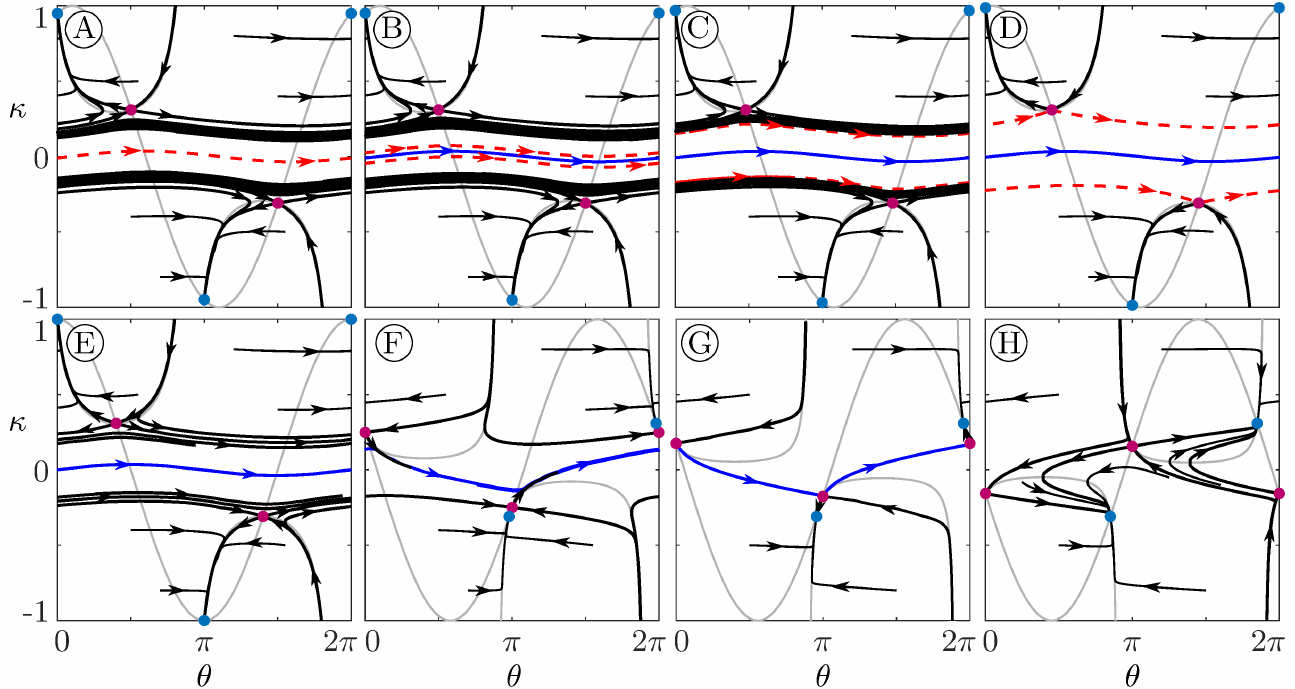}
\caption{Phase portraits for two-dimensional system~\eqref{eq:2Cl_red_system_theta}--\eqref{eq:2Cl_red_system_kappa}.
The graphics show the two classes of asymptotic states that
are equilibria (colored nodes) and periodic solutions (colored lines).
The stability properties of the individual asymptotic states are indicated
by the coloring where the blue refers to stable and the red (dashed)
to unstable states. In addition, several trajectories are plotted
in black including those close to the stable and unstable manifold
of the equilibria. The nullclines are displayed as gray lines. For
the different panels parameter $\beta$ is varied as shown in Fig.~\ref{fig:map_regimes}:
(A) $\beta=-0.601\pi$; (B) $\beta=-0.599\pi$; (C) $\beta=-0.58\pi$;
(D) $\beta=-0.5515\pi$; (E) $\beta=-0.5\pi$; (F) $\beta=-0.08\pi$;
(G) $\beta=-0.0563\pi$; and (H) $\beta=0.05\pi$. The other parameters
are $\alpha=0.1\pi$ and $\epsilon=0.01$. \label{fig:emergence_solitaries} }
\end{figure}

In the following, we study the structure of the phase space of~\eqref{eq:2Cl_red_system_theta}--\eqref{eq:2Cl_red_system_kappa}
for fixed $\alpha=0.1\pi$ and different values of parameter $\beta$.
Several bifurcation scenarios are discovered which give rise to the
birth and stability changes of the solitary states. We observe how
the stable solitary state emerges in a subcritical pitchfork bifurcation
of periodic orbits and disappears in a homoclinic bifurcation with
increasing $\beta$. Note that solitary states are given by periodic
solutions, where the phase difference $\theta(t)$ rotates. Equilibria
of this system describe one-cluster states. Figure~\ref{fig:emergence_solitaries}
shows several characteristic phase portraits of~\eqref{eq:2Cl_red_system_theta}--\eqref{eq:2Cl_red_system_kappa}
illustrating the bifurcation scenarios with increasing $\beta$ (from
(A) to (H)), see also Fig.~\ref{fig:map_regimes}. 

In Fig.~\ref{fig:emergence_solitaries}(A), we observe four equilibria
which correspond to certain one-cluster solution. The stable equilibria
at $\theta=0$ and $\theta=\pi$ correspond to in-phase synchronous
and antipodal where $a_{1}=\pi$ and $a_{i\ne1}=0$, respectively.
The other two saddle equilibria correspond to the special class of
double antipodal states~\cite{BER19} and describe therefore phase-cluster
similar to those described in~\cite{MAI14,TEI19}. While these equilibria
can be stable for the reduced system~\eqref{eq:2Cl_red_system_theta}--\eqref{eq:2Cl_red_system_kappa},
they are always unstable for~\eqref{eqn:AKS_Mdl_phi}--\eqref{eqn:AKS_Mdl_kappa}
in case of global coupling~\cite{BER19a}. Additionally, to these
four equilibria we find an unstable periodic orbit which corresponds
to an unstable solitary state.

With increasing $\beta$, we observe a subcritical pitchfork bifurcation
of periodic orbits at $\alpha+\beta=-\pi/2$ in which the unstable
periodic orbit is stabilized and two additional periodic orbits are
created. Figure~\ref{fig:emergence_solitaries}(B) shows the phase
portrait directly after the pitchfork bifurcation. Therefore, we conclude
that there exist three solitary states, two of which are unstable
and one stable. It is worth to remark that the stability for the reduced
system is only necessary but not sufficient to be a stable asymptotic
state for the network~\eqref{eqn:AKS_Mdl_phi}--\eqref{eqn:AKS_Mdl_kappa}.

By increasing $\beta$ even further the basin of attraction of the
stable periodic orbit increases and its boundaries are given by the
unstable periodic orbits, see~Fig.~\ref{fig:emergence_solitaries}(C).
For $\beta=-0.5515\pi$, the trajectories of the unstable solitary
states merge with the equilibria and become homoclinic orbits of the
saddle equilibria (Fig.~\ref{fig:emergence_solitaries}(D)). The
phase portrait after this homoclinic bifurcation is shown in Fig.~\ref{fig:emergence_solitaries}(E).

After the homoclinic bifurcation, with increasing $\beta$, the equilibria
are moving towards each other in phase space and exchange their stability
in a transcritical bifurcation. This can seen analytically by considering
the determining equations for the equilibria $(\theta,\kappa)$ of~\eqref{eq:2Cl_red_system_theta}--\eqref{eq:2Cl_red_system_kappa}
\begin{align}
0&=\cos(\alpha+\beta)-\cos(2\theta+\alpha+\beta),\label{eq:equilibria_red_system}\\
\kappa&=-\sin(\theta+\beta).\nonumber 
\end{align}
In general, equation~\eqref{eq:equilibria_red_system} possesses
two solutions for $2\theta$. At $\alpha+\beta=0,\pi$, however, these
two solutions coincide which describes the point of the transcritical
bifurcation. The stability of the equilibria can be further computed
by considering the two-dimensional system linearized around the equilibria.
In a more general setup this has be done Sec.~\ref{sec:StabilityAnalysis}.
Figure~\ref{fig:emergence_solitaries}(F) displays the phase portrait
after the transcritical bifurcation. Remember that although the double
antipodal clusters are stable for the reduced system, they are always
unstable for the full system~\cite{BER19a}.

In Figure~\ref{fig:emergence_solitaries}(G,H) another homoclinic
bifurcation is presented in which the stable solitary state becomes
a homoclinic orbit of the in-phase and antipodal cluster. The phase
portrait close to the homoclinic bifurcation is presented in Fig.~\ref{fig:emergence_solitaries}(H).
After the homoclinic bifurcation the phase space is divided into the
basins of attraction of the two double antipodal states and no more
solitary state exist.


\section{Adaptive networks with global base topology versus ring base topology: the differences}

Adaptive networks of coupled phase oscillators have been extensively studied on an all-to-all base structure~\cite{AOK09,AOK11,KAS17,BER19,BER19a}. This article extends the previous work towards more complex base topologies by considering a nonlocal ring base topology on which adaptation takes place. In this section we briefly summarize the main differences resulting from the different base topologies.

For the global base topology, all links between the nodes with the same frequency become active leading to the all-to-all structures within each cluster. Therefore, strongly connected components can be equivalently described by the frequency synchronization of nodes. 
In contrast to this, for ring networks (also more complex base structures), 
the frequency synchronization does not necessarily imply connectivity, 
see Fig.~\ref{fig:subnetworks}. 
As a result, we have adapted the definition of the frequency cluster on a complex base topology as a connected subnetwork with frequency synchronized oscillators. 

Another effect induced by the ring base topology concerns the hierarchical ordering of cluster sizes. For global base structures is has been found that a sufficiently large difference of the cluster sizes is necessary for the appearance of multicluster states. In case of a ring structure, the hierarchy is not necessary anymore. In figure~\ref{fig:multicl}(G--I), we present a five-cluster states that possesses two clusters each of size $7$ and additionally two clusters each of size $2$. For solitary states this nonhierarchical clustering implies that on a ring structure there can be several solitary nodes (Fig.~\ref{fig:solitaries}(D--F)). A simple explanation for the appearance of the clusters of a similar size is based on the fact that such clusters can be uncoupled in the base coupling structure and, hence, not synchronized. In contrast, in networks with the global base structure, similar clusters tend to be synchronized and merge into one larger cluster. 

Regarding the stability of rotating-wave states another striking difference between global and ring base topology is observed. Here the differences are twofold. For a global base topology rotating-waves constitute a $N-2$-dimensional family of solutions with the same collective frequency. On a nonlocal ring, this invariant family is not present anymore and all rotating-wave are different from each other including their frequencies. The same holds true for their stability. While the stability features of all rotating-waves agree on global structures, the stability properties depend crucially on the wavenumber (see Fig.~\ref{fig:1cl_stab}(A--C)).

\section{Conclusion}\label{sec:conclusion}
In summary, a model of adaptively coupled identical phase oscillators on a nonlocal ring has been studied. Various frequency synchronized states are observed including one-cluster, multicluster, and solitary states. Those states are similar to those found for a global base topology~\cite{KAS17,BER19}. However, to account for the complex base topology, we introduce a new definition of one-clusters by means of connected induced subnetworks. This definition allows furthermore to distinguish between multicluster and solitary states in a more strict way than it was done before.

Since one-cluster states form building blocks for multicluster states \cite{BER19a}, we have first investigated the existence and stability properties of one-cluster states. Here, we have introduced a novel type of phase-locked states for complex networks, namely local splay states, and have shown that this class of states is nonempty for any nonlocal ring base topology. In particular, we have proved that rotating-wave as well as antipodal states are always phase-locked solutions. Compared with the case of a global base topology, the different clusters of local splay type on a nonlocal ring structure can possess different collective frequencies. In addition, we have proved that local splay cluster are always global splay cluster. This statement relates, therefore, local with global (with respect to ''spatial'' structures in the network) properties of solutions.

The stability features of rotating-wave states have been studied numerically and analytically. The comparison of both approaches results in a very good agreement. Due to the analytic findings for rotating-wave states on a nonlocal ring, we are able to describe their stability depending on the coupling range $P$ and the wavenumber $k$. The limiting case of global coupling, i.e. $P=N/2$, is shown to be in agreement with the results presented in~\cite{BER19}.

An interesting feature of the system's behavior are solitary states. They have been previously found to emerge in the Kuramoto-Sakaguchi model with inertia~\cite{JAR18}. In this article, we show that solitary states are born in a homoclinic bifurcation and can be (de)stabilized in a pitchfork bifurcation of periodic orbits. In order to show this, a two-dimensional effective model is derived governing the dynamics of  solitary states. In contrast to the Kuramoto-Sakaguchi model with inertia, we observe a much more complicated bifurcation behavior. In particular, three different solitary states are created due to two individual homoclinic bifurcations. Two of these three solitary states, however, are unstable and bifurcate together with stable solitary states in a subcritical pitchfork bifurcation of periodic orbits.

Our results highlight the delicate interplay between adaptivity and complexity of the network structure. Since this interplay has been rarely investigated from the mathematical viewpoint, so far, this article raises many questions for future research which could be conducted for different network structures beyond nonlocal rings, other dynamical models for the local dynamics, nonidentical units or different adaptation rules.

\begin{acknowledgement}
	The authors acknowledge financial support by the Deutsche Forschungsgemeinschaft (DFG) (German Research Foundation)--Project Nos. 411803875 and 308748074, and by DAAD within the RISE programme.
\end{acknowledgement}

%

\appendixtitleon
\appendixtitletocon

\begin{appendices}
\section{From local to global order parameter}\label{app:Local2Global}
From expression \eqref{eq:const} we derived two types of one-cluster solution namely antipodal or local splay states. In the following we derive a remarkable relation between local and global properties on a nonlocal ring which is: If a cluster is of local splay type, the cluster is also of global splay type.

In order to show this, we rewrite the sum $\sum_{i=1}^{N} Z^{(2)}_{i} $ in two ways. Firstly, using the pure definition of the local order parameter~\eqref{eqn:local_order}:

\begin{align} \label{from_def}
    \sum_{i=1}^{N} Z^{(2)}_{i}(\bm{\chi}) = \frac{1}{2P}\sum_{i,j=1}^{N} a_{ij} e^{\mathrm{i} 2 \chi_{j}}= \sum_{j=1}^{N} e^{\mathrm{i} 2 \chi_{j}}= N Z^{(2)}(\bm{\chi}).
\end{align}

Secondly, the sum can be rewritten using the definition of a local splay type cluster with $R^{(2)}_{i}(\bm{\chi})=R^{(2)}_c(\bm{\chi})$ and $2 \chi_{i} = \vartheta^{(2)}_{i}$. Then

\begin{align}\label{from_const}
    \sum_{i=1}^{N} Z^{(2)}_{i}(\bm{\chi})  = \sum_{i=1}^{N} R^{(2)}_c e^{\mathrm{i} \vartheta^{(2)}_{i}} = R^{(2)}_c(\bm{\chi}) \sum_{i=1}^{N}  e^{\mathrm{i} 2 \chi_{i}}  = N R^{(2)}_c (\bm{\chi}) Z^{(2)}(\bm{\chi})
\end{align}
By equating~\eqref{from_def} and~\eqref{from_const} we obtain:

\begin{align*}
    (1 - R^{(2)}_c(\bm{\chi})) Z^{(2)}(\bm{\chi}) = 0
\end{align*}

The latter equation yields $R^{(2)}(\bm{\chi})=0$ for all local splay type clusters, since $R^{(2)}_c(\bm{\chi}) < 1$ by definition.
\section{Stability analysis for one-cluster states}\label{app:StabOneCluster}
First note that the set of equations~\eqref{eq:Linearized_phi}--\eqref{eq:Linearized_kappa} can be brought into the following block form
\begin{align}\label{eq:LinearisationBlockForm}
	\frac{\mathrm{d}}{\mathrm{d}t}
	\begin{pmatrix}
		\delta\bm{\phi}\\
		\delta\kappa
	\end{pmatrix}
	=
	\begin{pmatrix}
		M & B\\
		C & -\epsilon\mathbb{I}_{N^2}
	\end{pmatrix} 
	\begin{pmatrix}
		\delta\bm{\phi}\\
		\delta\kappa
	\end{pmatrix}
\end{align}
where $\left(\delta\bm{\phi}\right)^{T}=\left(\delta\phi_{1},\dots,\delta\phi_{N}\right)$, $\left(\delta\kappa\right)^{T}=\left(\delta\kappa_{11},\dots,\delta\kappa_{1N},\delta\kappa_{21},\dots,\delta\kappa_{NN}\right)$, \- $B=\begin{pmatrix}B_1 & \cdots & B_N
\end{pmatrix}$, $C=\begin{pmatrix}C_1\\ \vdots \\C_N
& \\
\end{pmatrix}$, and $M$, $B_n$, $C_n$ are $N\times N$ matrices with $n=1,\dots,N$. The elements of the block matrices read
\begin{align*}
	m_{ij} & =\begin{cases}
		\!\begin{aligned}
			-\frac{\sigma}{2}\sin(\alpha-\beta)\sum_{k=1}^N a_{ik}-\sigma a_{ii}\sin(\beta)\cos(\alpha) \\
			+\frac{\sigma}{2}\sum_{k=1}^{N}a_{ik}\sin(2(a_{i}-a_{k})+\alpha+\beta),
		\end{aligned} 
		& i=j\\
		\frac{\sigma a_{ij}}{2}\left(\sin(\alpha-\beta)-\sin(2(a_{i}-a_{j})+\alpha+\beta)\right), & i\ne j
	\end{cases}\\
	& =\begin{cases}
		\!\begin{aligned}
		-\frac{\sigma}{2}\left(\sin(\alpha-\beta)+\Im(e^{-\mathrm{i}(2a_i+\alpha+\beta)}Z^{(2)}_i(\bm{a}))\right)\sum_{k=1}^N a_{ik}-\sigma a_{ii}\sin(\beta)\cos(\alpha)
		\end{aligned} 
		& i=j\\
		\frac{\sigma a_{ij}}{2}\left(\sin(\alpha-\beta)-\sin(2(a_{i}-a_{j})+\alpha+\beta)\right), & i\ne j
	\end{cases}\\
	b_{ij;n} & =\begin{cases}
		-\sigma a_{nj}\sin(a_{n}-a_{j}+\alpha), & i=n\\
		0, & \text{otherwise}
	\end{cases}\\
	c_{ij;n} & =\begin{cases}
		0, & j=n,i=j\\
		-\epsilon a_{ni} \cos(a_{n}-a_{i}+\beta), & j=n,i\ne j\\
		\epsilon a_{ni} \cos(a_{n}-a_{i}+\beta), & j\ne n,i=j\\
		0, & \text{otherwise}
	\end{cases},
\end{align*}
where $\sigma=1/2P$. Throughout this appendix we will make use of Schur's complement~\cite{LIE15} in order to simplify characteristic equations. In particular, any $m\times m$ matrix $M$ in the $2\times2$ block form can be written as
\begin{align}\label{eq:SchurComplement}
	L =\begin{pmatrix}M & B\\
		C & E
	\end{pmatrix}=\begin{pmatrix}\mathbb{I}_{p} & BE^{-1}\\
		0 & \mathbb{I}_{q}
	\end{pmatrix}\begin{pmatrix}M-BE^{-1}C & 0\\
		0 & E
	\end{pmatrix}\begin{pmatrix}\mathbb{I}_{p} & 0\\
		E^{-1}C & \mathbb{I}_{q}
	\end{pmatrix}
\end{align}
where $M$ is a $p\times p$ matrix and $E$ is an invertible $q\times q$ matrix. The matrix $M-BE^{-1}C$ is called Schur's complement. A simple formula for the determinant of $L$ can be derived with this decomposition in~\eqref{eq:SchurComplement}
\begin{align*}
	\det(L) & =\det(M-BD^{-1}C)\cdot\det(E).
\end{align*}
This result is important for the subsequent stability analysis. Note that in the following an asterisk indicates the complex conjugate.	Suppose we have a phase locked state with phases $a_i\in[0,2\pi)$. Then, the solution for the characteristic equations corresponding to the linearized system~\eqref{eq:Linearized_phi}--\eqref{eq:Linearized_kappa} are given by $\lambda=-\epsilon$ with multiplicity $N^2-N$ and by the solution of the following set of equations
\begin{align}\label{eq:SchurDeterminant}
	\det\left(\left(M-\lambda\mathbb{I}_{N}\right)\left(\epsilon+\lambda\right)+BC\right) = 0.
\end{align}
The second term	$D:=BC$ of Schur's complement are element wise given by
	\begin{align*}
		d_{ij} &=-\frac{\epsilon\sigma}{2} a_{ij}\left(\sin(\alpha-\beta)+\sin(2(a_{i}-a_{j})+\alpha+\beta)\right)
	\end{align*}
	if $i\ne j$ and
	\begin{align*}
		d_{ii} 
		&= \frac{\epsilon\sigma}{2}\left(\sin(\alpha-\beta)-\Im(e^{-\mathrm{i}(2a_i+\alpha+\beta)}Z^{(2k)}_i(\bm{a}))\right)\sum_{k=1}^N a_{ik} -\epsilon\sigma\sin(\alpha)\cos(\beta)
	\end{align*}
	Consider that $a_i(k)=ik\frac{2\pi}{N}$ and the base topology has constant row sum $\rho\in\mathbb{N}$, i.e. $\sum_{k=1}^N a_{ik}=\rho$, then the matrix in~\eqref{eq:SchurDeterminant} becomes circulant. For a ring structure, as considered in this paper, we have $\rho=2P$. Hence it can be diagonalized using the $N$ eigenvectors $\zeta_l = \exp(\mathrm{i}2\pi l/N)=\exp(\mathrm{i}a_l(1))$ and the eigenvalues $\mu_l(\lambda)$ are
	\begin{multline*}
		\mu_l = (\left(m_{NN}-\lambda\right)\left(\epsilon+\lambda\right)+d_{NN}) + \sum_{j=1}^{N-1} (m_{Nj}(\epsilon+\lambda)+d_{Nj})\zeta_l^j \\
		= -\lambda^2 + \frac{\rho\sigma}{2}\lambda\left[\frac{\mathrm{i}}{2}\left(e^{\mathrm{i}(\alpha+\beta)} \left(Z_N^{(l-2k)}-Z_N^{(2k)}\right)-e^{-\mathrm{i}(\alpha+\beta)} \left(Z_N^{(2k+l)}-Z_N^{(2k)}\right)\right)\right.\\
		\left.+(Z^{(l)}_N - 1)\sin(\alpha-\beta)-\frac{2\epsilon}{\rho\sigma}\right]\\
		+\epsilon\sigma\rho\frac{\mathrm{i}}{2}\left(e^{\mathrm{i}(\alpha+\beta)} \left(Z_N^{(l-2k)}-Z_N^{(2k)}\right)-e^{-\mathrm{i}(\alpha+\beta)} \left(Z_N^{(2k+l)}-Z_N^{(2k)}\right)\right).
	\end{multline*}
	Here, we use the shorthand form $Z_N^{(n)}=Z_N^{(n)}(1)$. If any of the $\mu_l$ vanishes, the determinant in~\eqref{eq:SchurDeterminant} vanishes, as well. Thus, $\mu_l(\lambda)=0$ is giving the quadratic equation which determine the Lyapunov spectrum of the linearized system~\eqref{eq:Linearized_phi}--\eqref{eq:Linearized_kappa} around a phase-locked solution and with a base topology, given by the adjacency matrix $A$, having constant in-degree. Note that with the result in Sec.~\ref{sec:equlibria}, the complex order parameter can be further simplified
	\begin{align}
        Z_N^{(n)}=R_N^{(n)}= \frac{1}{P}\left(\sum_{j=1}^{P}\cos(nj\frac{2\pi}{N})\right).
    \end{align}
    Using the latter equation and well-known trigonometric equations, the following relations are derived:
    \begin{align*}
        Z_N^{(l-2k)}-Z_N^{(2k)} &= \frac{1}{P}\left(\sum_{j=1}^{P}\left(\cos((l-2k)j\frac{2\pi}{N})-\cos(2kj\frac{2\pi}{N})\right)\right),\\
        &=-\frac{2}{P}\left(\sum_{j=1}^{P}\sin(lj\frac{\pi}{N})\sin((l-4k)j\frac{\pi}{N})\right),\\
        &=-\frac{2}{P}\left(\sum_{j=1}^{P}\sin(lj\frac{\pi}{N})\left(\sin(lj\frac{\pi}{N})\cos(4kj\frac{\pi}{N})-\cos(lj\frac{\pi}{N})\sin(4kj\frac{\pi}{N})\right)\right),
    \end{align*}
    and analogously
    \begin{align*}
        Z_N^{(l+2k)}-Z_N^{(2k)} 
        &=-\frac{2}{P}\left(\sum_{j=1}^{P}\sin(lj\frac{\pi}{N})\left(\sin(lj\frac{\pi}{N})\cos(4kj\frac{\pi}{N})+\cos(lj\frac{\pi}{N})\sin(4kj\frac{\pi}{N})\right)\right).
    \end{align*}
    Combining these relations, we find
    \begin{multline}\label{eq:def_Lfunction}
        L(\alpha,\beta,l,k)=\frac{\mathrm{i}}{2}\left(e^{\mathrm{i}(\alpha+\beta)} \left(Z_N^{(l-2k)}-Z_N^{(2k)}\right)-e^{-\mathrm{i}(\alpha+\beta)} \left(Z_N^{(2k+l)}-Z_N^{(2k)}\right)\right)\\
        = \frac{2\sin(\alpha+\beta)}{P}\sum_{j=1}^{P}\sin^2(lj\frac{\pi}{N})\cos(4kj\frac{\pi}{N}) \\+ \mathrm{i}\frac{2\cos(\alpha+\beta)}{P}\sum_{j=1}^{P}\sin(lj\frac{\pi}{N})\cos(lj\frac{\pi}{N})\sin(4kj\frac{\pi}{N}).
    \end{multline}
\end{appendices}
\end{document}